# Oxygen Partial Pressure during Pulsed Laser Deposition: Deterministic Role on Thermodynamic Stability of Atomic Termination Sequence at SrRuO$_3$/BaTiO$_3$ Interface

Yeong Jae Shin[†,‡], Lingfei Wang[†,‡,*], Yoonkoo Kim[§], Ho-Hyun Nahm[†,‡], Daesu Lee[†,‡], Jeong Rae Kim[†,‡], Sang Mo Yang[∥], Jong-Gul Yoon[#], Jin-Seok Chung[⊥], Miyoung Kim[§], Seo Hyoung Chang[¶,*], Tae Won Noh[†,‡,*]

[†]Center for Correlated Electron Systems, Institute for Basic Science (IBS), Seoul 08826, Republic of Korea

[‡]Department of Physics and Astronomy, Seoul National University, Seoul 08826, Republic of Korea

[§]Department of Materials Science and Engineering and Research Institute of Advanced Materials, Seoul National University, Seoul 08826, Republic of Korea

[∥]Department of Physics, Sookmyung Women's University, Seoul 04310, Republic of Korea

[#]Department of Physics, University of Suwon, Hwaseong, Gyunggi-do 18323, Republic of Korea

[⊥]Department of Physics, Soongsil University, Seoul 06978, Republic of Korea

[¶] Department of Physics, Chung-Ang University, Seoul 06974, Republic of Korea

**ABSTRACT**

With recent trends on miniaturizing oxide-based devices, the need for atomic-scale control of surface/interface structures by pulsed laser deposition (PLD) has increased. In particular, realizing uniform atomic termination at the surface/interface is highly desirable. However, a lack of understanding on the surface formation mechanism in PLD has limited a deliberate control of surface/interface atomic stacking sequences. Here, taking the prototypical $SrRuO_3/BaTiO_3/SrRuO_3$ (SRO/BTO/SRO) heterostructure as a model system, we investigated the formation of different interfacial termination sequences ($BaO-RuO_2$ or $TiO_2-SrO$) with oxygen partial pressure ($P_{O2}$) during PLD. We found that a uniform $SrO-TiO_2$ termination sequence at the SRO/BTO interface can be achieved by lowering the $P_{O2}$ to 5 mTorr, regardless of the total background gas pressure ($P_{total}$), growth mode, or growth rate. Our results indicate that the thermodynamic stability of the BTO surface at the low-energy kinetics stage of PLD can play an important role in surface/interface termination formation. This work paves the way for realizing termination engineering in functional oxide heterostructures.

**KEYWORDS**

Ferroelectric, interface engineering, oxide heterostructure, pulsed laser deposition, thermodynamic surface stability

**INTRODUCTION**

Perovskite oxide-based heterostructures have been investigated extensively for decades due to the intriguing physical properties and great potential in device applications.[1–5] When reducing the film thickness down to the nanometer scale, the detailed atomic arrangements at surfaces and heterointerfaces become crucial in exploring emergent phenomena and determining various functionalities, including electric transport, magnetism, and ferroelectricity.[6–11] For instance, recent studies on $SrRuO_3/BaTiO_3/SrRuO_3$ (SRO/BTO/SRO) ferroelectric capacitors showed that the atomically sharp and symmetric SRO/BTO interfaces can stabilize the ferroelectricity of BTO down to the theoretical thickness limit.[12] To obtain such atomically sharp interfaces and uniform termination sequences in $ABO_3$ perovskite-based heterostructures, it is necessary to understand and control the stacking sequence of atomic layers at the half unit-cell level (e.g., AO or $BO_2$ layer).[9,13,14]

Pulsed laser deposition (PLD) has been considered as one of the most powerful growth techniques for realizing high-quality oxide heterostructures. Owing to high-energy laser ablation, PLD enables effective conveyance of the target material's stoichiometry to the substrate.[15,16] Combined with reflection high-energy electron diffraction (RHEED), PLD can also be used to monitor layer-by-layer growth and realize unit-cell level control of perovskite oxides.[17,18] However, during PLD growth, the expected stacking sequences according to the layer-by-layer growth mode can be distinct from actual atomic arrangements. For instance, during the growth of SRO on a $TiO_2$-terminated $SrTiO_3(001)$ [STO(001)] substrate, reversal of the surface termination from $RuO_2$ to SrO occurs within a few unit cells due to the highly volatile nature of $RuO_2$.[19,20] To achieve the desired termination sequence, the sequential growth of an additional binary oxide monolayer has been used.[6,21–23] However, this method easily induces surface roughening or stacking faults except for a few constituents (i.e., SrO). Very recently, studies on the growth of manganite and titanate films have shown that the atomic stacking sequences at surfaces/interfaces are strongly dependent on the growth parameters, specifically the background atmosphere during PLD.[12,24] Although these works strongly imply a potential way to realize in situ control of the surface/interface termination sequence, the underlying mechanisms have yet to be resolved.

To elucidate the role of background atmosphere, we break down the PLD processes broadly into three sequential stages.[15,16] i) The laser pulse is absorbed by the target surface and a plasma plume containing the ablated materials is generated. ii) The plume starts to expand and bombard the substrate with ionic species. The plume dynamics is governed by the ionic species with high kinetic energy up to several hundred eV.[25] iii) Nucleation is triggered by supersaturated materials on the substrate and the film starts to grow with the diffusion of adatoms. The kinetic energy of adatoms on the film surface is on the scale of several eV.[16,26] The background atmosphere can strongly affect stages ii) and iii), which are denoted as high- and low-energy kinetics stages, respectively.

In the high-energy kinetics stage, background gas can affect the film stoichiometry because the detailed scattering of each cation depends strongly on the gas composition and pressure.[27–30] In addition, the background gas can significantly reduce the kinetic energy of ionic species arriving at the substrate by one or two orders of magnitude.[26] Thus, it can affect the density and shape of nucleation sites as well as the growth rate.[27,31] During the low-energy kinetics stage, the adatoms lose their kinetic energies until covalent bonds form. Surface structuring is affected significantly by the energy landscape of the film surface. The background atmosphere can act as an important thermodynamic parameter to determine the surface energy landscape and thus the ultimate surface structure.[32] Therefore, understanding the role of the growth parameters at each PLD stage is of particular importance in controlling the ultimate surface/interface termination sequence.

In this study, using the SRO/BTO/SRO heterostructure as a model system, we investigated the evolution of an interfacial termination sequence with oxygen partial pressure ($P_{O2}$) during PLD. The interfacial termination sequence in SRO/BTO/SRO heterostructures is sensitive to the growth atmosphere and is crucial for stabilizing the ferroelectricity of BTO.[12,33,34] To elucidate the mechanism of termination sequence variation with $P_{O2}$, we monitored the growth of a BTO layer under background atmospheres with different gas compositions. Systematic structural characterizations on the as-grown films were performed; the results showed that the termination sequence at the SRO/BTO interface is determined by $P_{O2}$ only, regardless of total background gas pressure ($P_{total}$), growth mode, or growth rate. Density functional theory (DFT) calculations and phenomenological modeling indicated that the thermodynamic stability of the BTO surfaces (i.e., BaO- or $TiO_2$-terminated surface) during the low-energy kinetics stage is highly dependent on $P_{O2}$. Based on the results, we suggest that the difference in the thermodynamic surface stabilities is the main driving force of the $P_{O2}$-dependent termination sequence at the SRO/BTO interface.

**RESULTS AND DISCUSSION**

All of the SRO/BTO/SRO heterostructures studied herein were grown *via* PLD on atomically smooth $TiO_2$-terminated STO(001) substrates.[12,35] During the deposition process, the substrate temperature was maintained at 1000 K. The SRO top and bottom electrodes were grown under pure oxygen atmosphere with $P_{O2}$ = 100 mTorr. The BTO layers were grown under various background atmospheres. Here, we present the results from three representative cases: 1) a pure oxygen atmosphere with $P_{total}$ = $P_{O2}$ = 5 mTorr, 2) a pure oxygen atmosphere with $P_{total}$ = $P_{O2}$ =

150 mTorr, and 3) an argon and oxygen (O$_2$/Ar) mixed atmosphere with $P_{O2}$ = 5 mTorr and $P_{total}$ = 150 mTorr.

Figure 1 shows the photographic images of a laser-ablated plume and a time-dependent intensity profile of an RHEED specular spot during BTO growth (RHEED intensity profiles during bottom- and top-SRO growths can be found in Figure S1 in Supporting Information). For the samples grown at $P_{O2}$ = 150 and 5 mTorr, the intensity profiles clearly exhibit four oscillations, indicating standard layer-by-layer growth (Figure 1b,d, respectively). For the sample grown in an O$_2$/Ar mixed atmosphere, the layer-by-layer growth persisted up three oscillations of RHEED intensity (solid line in Figure 1f). With further growth, the RHEED intensity decreased significantly (dashed line in Figure 1f). This behavior implies a growth mode transition from layer-by-layer growth to Stranski–Krastinov type (layer-plus-island) growth.[36]

The high epitaxial quality of the SRO/BTO/SRO heterostructures was confirmed *via* atomic force microscopy (AFM) and X-ray diffraction (XRD). Figure 2a–c shows AFM topographic images of SRO/BTO/SRO heterostructures grown under the three background atmospheres. Please note that to avoid possible step-bunching of SRO layer, we intentionally choose the STO(001) substrates with terraces larger than 300 nm. Even though the larger step-width caused meandering of step-edge by coalescence of 2D islands with advancing steps during SRO grwoths,[37] all of the samples exhibited atomically flat surfaces with uniform 1-unit cell (uc)-high terraces.[19,37,38] The XRD $2\theta$-$\omega$ scans shown in Figure 2d are almost identical. The observed high-intensity SRO(220) peaks and clear Laue fringes signify sharp heterointerfaces. Note that the {00$l$} peaks from the ultrathin BTO layers were too weak to be detected by XRD.

Despite nearly identical features in macroscopic structural characterizations, scanning transmission electron microscopy (STEM) results revealed distinct interfacial termination sequences in SRO/BTO/SRO heterostructures, depending on the background atmosphere. Figure 3a shows a high-angle annular dark field (HAADF) image measured from the BTO sample grown under $P_{O2}$ = 150 mTorr; the bottom BTO/SRO interface exhibited a uniform $TiO_2$-SrO termination sequence, while the top SRO/BTO interface showed a rather complicated structure. Figure 3b shows the magnified HAADF image and intensity line profiles from a preselected region (marked as a dashed box in Figure 3a); the right side of the image exhibits four $TiO_2$ layers and three BaO layers, implying BTO thickness of 3.5 uc. On the left side of the image, the topmost $TiO_2$ layer is gradually converted into a $RuO_2$ layer, reducing BTO thickness to 3 uc (Figure 3a). This result demonstrates heterogeneous terminations with both SrO-$TiO_2$ and BaO-$RuO_2$ sequences at the top SRO/BTO interface, as shown in Figure 3c. In contrast, for the sample grown at $P_{O2}$ = 5 mTorr, the HAADF images (Figure 3d,e) showed uniform SrO-$TiO_2$ termination sequences for both the bottom and the top SRO/BTO interfaces (Figure 3f). The uniformity of the termination sequence was confirmed by surface X-ray scattering experiments.[12]

To clarify the $P_{O2}$-dependent interfacial termination sequences further, we grew BTO samples under an $O_2$/Ar mixed atmosphere and performed STEM measurements (fixed $P_{total}$ = 150 mTorr but lower $P_{O2}$ = 5 mTorr). The BTO sample grown up to three RHEED oscillations showed a uniform $TiO_2$-SrO termination sequence at both the top and the bottom SRO-BTO interfaces, as shown in Figure 4a–c. This symmetric interfacial termination was the same as that of the sample grown in pure $P_{O2}$ = 5 mTorr. For the samples with further BTO growth, as shown in Figure 4d–f, islands 1 uc in height appeared at the top interface, leading to a local thickness variation of 1 uc This spatial structural inhomogeneity was consistent with the decay evident in the RHEED intensity profile shown in Figure 1f. Despite these differences in interface structure, both the thinner (2.5 uc) and the thicker (3.5 uc) regions exhibited a symmetric, uniform $TiO_2$-SrO termination sequence. The heterogeneous termination structure was absent throughout the entire sample. Accordingly, the STEM results strongly suggest that the interfacial termination sequence is determined by $P_{O2}$ rather than $P_{total}$.

The observed interfacial termination sequences at the top SRO/BTO interface were inconsistent with the stacking sequence of BTO with the layer-by-layer growth scheme. The as-grown SRO bottom electrode has a uniformly SrO-terminated surface due to the highly volatile nature of the $RuO_2$ layer.[19] Accordingly, the unit-cell-by-unit-cell stacked BTO should have a BaO-terminated surface. The resultant top SRO/BTO interface was expected to show a uniform $RuO_2$-BaO termination sequence, which has never been observed in our samples. In addition, the number of BaO and $TiO_2$ layers observed in HAADF images were not same with that predicted by intensity profiles of RHEED specular spot. On this basis, we suggest that altering $P_{O2}$ could trigger atomic-scale structural modulations, such as selective decomposition of BaO or $TiO_2$ layer, on the BTO surface during growth, thus leading to changes in the termination sequence. The atomic-scale structural modulations can be triggered by either plume dynamics variation during the high-energy kinetics stage or changes in surface structuring during the low-energy kinetics stage. In the following part of this paper, we attempt to resolve the effects of varying $P_{O2}$ in terms of these PLD stages.

The changes in plasma plume dynamics with background atmosphere during the high-energy kinetics stage can be clearly manifested by the RHEED results in Figure 1. During the growth of the BTO sample under a pure oxygen atmosphere, the laser-ablated plume with $P_{O2}$ = 150 mTorr (Figure 1a) was larger and brighter than that with $P_{O2}$ = 5 mTorr (Figure 1c), implying a stronger interaction with the background atmosphere.[16] Moreover, because the high-pressure environment slows down the high energetic ions inside the plume, the growth rate in the $P_{O2}$ = 150 mTorr case (Figure 1b) was also lower than that in the $P_{O2}$ = 5 mTorr case (Figure 1d). To differentiate the effects of $P_{O2}$ from those of $P_{total}$, we also monitored the growth under an O$_2$/Ar mixed atmosphere, of which $P_{total}$ = 150 mTorr while $P_{O2}$ = 5 mTorr. As shown in Figure 1e,f, both the plume characteristics (Figure 1e) and the growth rate (Figure 1f) under the O$_2$/Ar mixed atmosphere were similar to those in the $P_{O2}$ = 150 mTorr case, implying similar plume dynamics. Comparing the RHEED results in the three cases, the plume dynamics were mainly determined by $P_{total}$ rather than $P_{O2}$. Nevertheless, STEM results demonstrated that the interfacial termination sequence is determined by $P_{O2}$ only. On this basis, we assert that the plume dynamics-related structural changes cannot explain the different termination sequences at the top SRO/BTO interfaces.

We also confirmed that the variation in chemical stoichiometry, which could also be sensitive to plume dynamics, cannot explain the $P_{O2}$-dependent termination sequences. We investigated the cation stoichiometry of BTO films by STEM-based energy-dispersive spectroscopy (EDS). Figure 5a shows a Ti/Ba cation ratio of 20-nm-thick BTO films grown under pure oxygen atmosphere with various $P_{O2}$, as well as in an O$_2$/Ar mixed atmosphere. For all of the films, the Ba/Ti ratios were nearly identical within the experimental error. We also investigated the oxygen stoichiometry by selected-area electron energy-loss spectroscopy (EELS).[13,39] Figure 5b shows an HAADF image of a SRO/BTO/SRO/STO(001) heterostructure grown at $P_{O2}$ = 5 mTorr. The EELS values at Ti-L and O-K core edges were measured from preselected areas in the 2.5 uc BTO layer and the STO substrate (open circles in Figure 5b). In Figure 5c, the EELS curves of BTO at the Ti-L (O-K) core edge exhibited four (three) well-defined peaks. The shapes and positions of these peaks were close to those of the stoichiometric STO substrate.[39] This implies that the oxygen deficiency was minimized, even for our sample grown at lower $P_{O2}$ = 5 mTorr.[13,39] Within the resolution of our experimental set-up, the chemical stoichiometry variation of the BTO layer was negligible and did not contribute to the observed interface structural variations.

Based on the above results and discussions, we suggest that the termination sequence variation at the SRO/BTO top interface occurs during the low-energy kinetics stage. During this stage, the surface structuring is accompanied by the kinetic energy loss of adatoms and the formation of covalent bonds. Depending on the chemical composition and bonding nature, the atomic structure of the film surface can have a complicated phase diagram.[40,41] Initially, the kinetic energy of adatoms is much higher than the energy barriers among different surface structures. Hence, after the adatoms lose kinetic energy, the ultimately formed surface structure should be the one with highest thermodynamic stability. The thermodynamic stability of the BTO surface can be described by the surface Gibbs free energy.[40,41] In the following part, we attempt to determine the Gibbs free energies of possible BTO surface structures via a phenomenological model based on a thermodynamic framework, together with DFT calculations.

For simplicity, here we consider the $TiO_2$- and BaO-terminated (1 × 1) bulk-like BTO surface only. The Gibbs free energies of these two surfaces are denoted as $\Omega^{TiO_2}$ and $\Omega^{BaO}$, respectively. According to the methodology described in the Supporting Information, the surface energy difference between these two surface structures $\Delta\Omega$ is given by:

$$\Delta\Omega = \Omega^{TiO_2} - \Omega^{BaO} = \phi^{TiO_2} - \phi^{BaO} + 2\Delta\mu_{Ba} + 2\Delta\mu_O, \qquad (1)$$

where the constants $\phi^{TiO_2}$ and $\phi^{BaO}$ are determined by the intrinsic structural and electronic properties of $TiO_2$ and BaO-terminated surfaces. We can obtain $\phi^{TiO_2} - \phi^{BaO} = 6.37$ eV from DFT calculations (Table S1 in Supporting Information). $\Delta\mu_O$ and $\Delta\mu_{Ba}$ are chemical potentials for each species with respect to their standard states at 0 K. $\Delta\mu_O$ is correlated with $P_{O2}$ according to the ideal gas relationship:

$$\Delta\mu_O(T, P_{O2}) = (E_{O_2}^{gas} + \Delta\mu_{O_2}(T, P_{O2}) + k_B T \ln(P_{O2}/P_0))/2, \tag{2}$$

where $T(T_0)$ is the temperature (enthalpy reference temperature, $T_0 = 298.15$ K), $P_0$ is the standard state pressure (760 mTorr), $E_{O_2}^{gas}$ is the total energy of the $O_2$ molecule, and $\Delta\mu_{O_2}(T, P_{O2})$ is the change in the oxygen chemical potential from the temperature-dependent enthalpy and entropy of the $O_2$ molecule.[42] $\Delta\mu_{Ba}$ is the only unknown parameter in eq 1. When $P_{O2}$ = 150 mTorr, the BTO surface exhibited mixed $TiO_2$ and BaO terminations, which indicates that the levels of thermodynamic stability of these two surfaces are similar. On this basis, we estimated $\Delta\mu_{Ba}$ = $-5.2$ eV to satisfy $\Delta\Omega = 0$ at $P_{O2}$ = 150 mTorr.

From eq 1 and eq 2, we can construct the relationship between $P_{O2}$ and $\Delta\Omega$. Figure 6 clearly shows a linear relationship between $\log(P_{O2})$ and $\Delta\Omega$ at $T$ = 1000 K. By reducing $P_{O2}$, $\Delta\Omega$ decreased monotonically, which made the $TiO_2$-terminated surface more stable. When $P_{O2}$ reached 5 mTorr, $\Delta\Omega$ decreased to $-280$ meV. The magnitude of this value is much larger than the thermal energy at the growth temperature (1000 K, corresponds to approximately 86 meV). Therefore, under a pure oxygen atmosphere of $P_{O2}$ = 5 mTorr, once the $TiO_2$-terminated BTO surface forms, it can be stabilized uniformly. For the $O_2$/Ar mixed atmosphere with the same $P_{O2}$, the additional Ar gas can significantly change the plume dynamics but preserve the surface stability due to its negligible reactivity. As a result, the top SRO/BTO interface, although exhibiting island characteristics, still exhibited a uniform SrO-$TiO_2$ termination sequence. At higher $P_{O2}$ near 150 mTorr, the magnitude of $\Delta\Omega$ became smaller than the thermal energy, leading to a mixed surface termination. Note that stabilizing a uniform BaO-terminated surface requires a high $P_{O2}$ over 300 mTorr, which is not possible for obtaining high-quality BTO films.

**CONCLUSIONS**

In summary, we revealed the deterministic role of $P_{O2}$ on the interfacial termination sequences in SRO/BTO/SRO heterostructures grown by PLD. Uniform SrO-TiO$_2$ interfacial termination was achieved by lowering the $P_{O2}$ to 5 mTorr. This effective termination control was driven by the different thermodynamic stabilities of TiO$_2$- and BaO-terminated surfaces, which evolve significantly with $P_{O2}$. Our work highlights the roles of the low-energy kinetics stage of PLD and thermodynamic parameters in the interface/surface structures of as-grown films, which has been largely overlooked. This scenario could be generalized to other oxide heterostructures for deliberately controlling the atomic structure of the interface, leading to the development and implementation of oxide-based electronic devices.

**EXPERIMENTAL SECTION**

**Film growth.** All of the SrRuO$_3$/BaTiO$_3$/SrRuO$_3$ (SRO/BTO/SRO) heterostructures were fabricated using pulsed laser deposition (PLD) with a KrF excimer laser (248 nm, COMPex pro, Coherent). Prior to deposition, SrTiO$_3$ (001) [STO(001)] substrates with low-angle-miscut (0.05–0.1°) were etched by buffered hydrofluoric acid and annealed at 1100 °C for 1 hour in air. This procedure allows the STO substrates to have an atomically smooth TiO$_2$-terminated surface with 1-unit cell (uc)-high terraces. During the deposition process, the substrate temperature was maintained at 1000 K and the laser fluence was set as 1.5 J/cm$^2$. The repetition rates of the laser were 3 Hz for SRO and 2 Hz for BTO. The SRO top and bottom electrodes were grown under pure oxygen atmosphere, with a partial pressure ($P_{O2}$) equal to 100 mTorr. The thickness of the electrodes was fixed at 20 nm for all samples used in the paper. During the entire PLD procedure, the surface of the film was monitored by reflection high-energy electron diffraction (RHEED). After deposition, the heterostructures were cooled at 20°C/min under $P_{O2}$ = 100 mTorr.

**AFM and XRD measurements.** The surface topographies of SRO/BTO/SRO heterostructures were investigated by atomic force microscopy (AFM, Asylum Cypher) in the non-contact tapping mode with a non-conductive tip (Nano Sensors). To determine the crystal structure of the SRO/BTO/SRO films, we performed X-ray diffraction (XRD) $2\theta$-$\omega$ scans using a commercial high-resolution X-ray diffractometer (Bruker AXS D8 with a Vantec line-detector) and Huber six-circle diffractometer at Sector 9C of the Pohang light source.

**Atomic structure.** The structure of the samples was investigated at the atomic scale via scanning transmission electron microscopy (STEM, JEM-ARM200F, JEOL) and coherent Bragg rod analysis (COBRA, Pohang Light Source). During STEM measurements, we obtained multiple cross-sectional high-angle annular dark field (HAADF) images of each sample; representative images are shown in Figures 3 and 4. Note that all of the images were viewed along the [001] zone axis.

**Calculation.** Density functional theory (DFT) calculations were performed using the Vienna Ab-initio Simulation Package (VASP). We constructed the supercell by stacking 4 uc BTO on top of 6 uc SRO. The BTO cell contains four $TiO_2$ layers and three (four) BaO layers for the $TiO_2$-terminated (BaO-terminated) surface.[40,41] Between the periodic slabs, 20 Å of vacuum was stacked along the [001] direction to prevent artificial interactions. To simplify the calculations, we neglected the ferroelectricity of BTO and considered the $TiO_2$- and BaO-terminated (1 × 1) bulk-like BTO surface only. The exchange-correlation function of the generalized gradient approximation (GGA) with the on-site $U$ (GGA+$U$) approach was utilized based on the Perdew–Burke–Ernzerhof (PBE) scheme.[43,44] For mimicking the magnetic and electronic properties of the itinerant ferromagnetic $SrRuO_3$ metal, an effective $U$ ($U_{eff}= U$-$J$) value of 1.0 eV for the $d$ orbital was chosen, which is consistent with other relevant DFT calculations.[45] Projector augmented-wave (PAW) pseudopotentials[46] were used along with a plane-wave basis set, with a kinetic energy cut-off of 400 eV and a 4 × 4 × 1 **k**-point mesh. Using the DFT calculations together with a phenomenological model based on the thermodynamic framework, we can determine the Gibbs free energy of BTO surfaces and their relative stabilities. The important parameters from DFT calculations and detailed methodology of phenomenological modeling are described in Supporting Information.

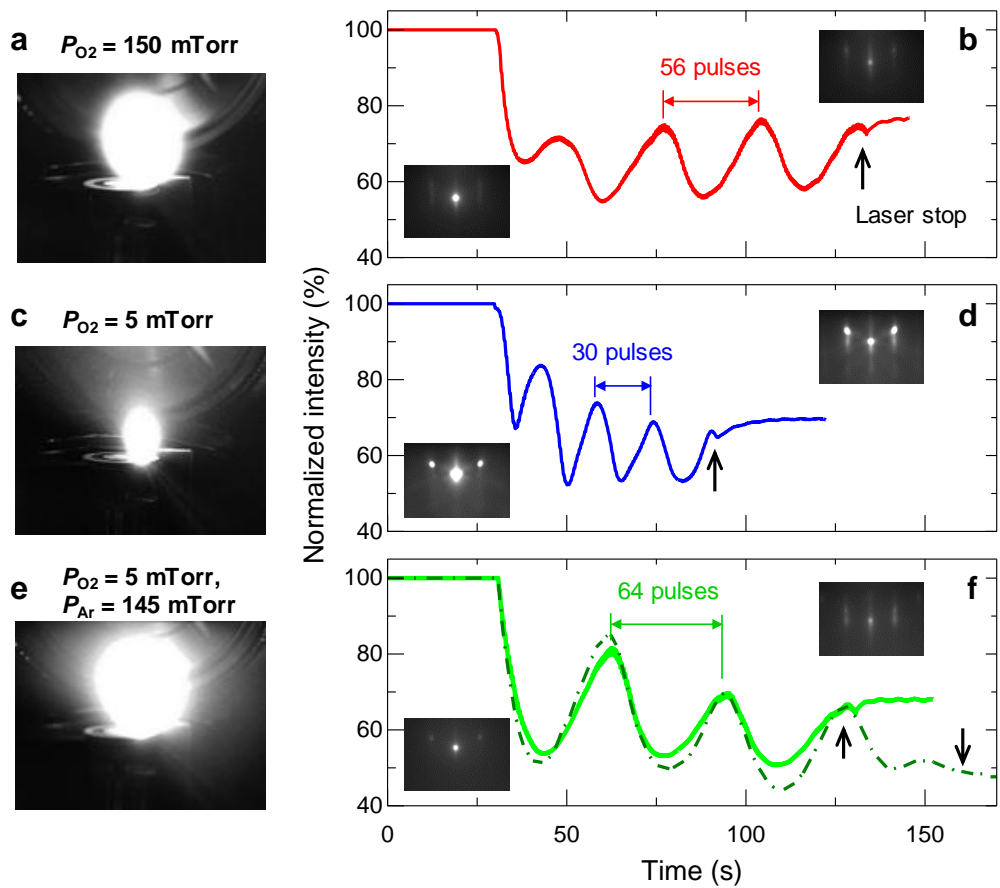

**Figure 1.** In situ monitoring of the growth of BaTiO$_3$ (BTO) layers. (a,c,e) Photographic images of the laser-ablated plume during BTO growth under (a) $P_{total} = P_{O2} = 150$ mTorr, (c) $P_{total} = P_{O2} = 5$ mTorr, and (e) O$_2$/Ar mixed atmosphere ($P_{total} = 150$ mTorr and $P_{O2} = 5$ mTorr). (b,d,f) Time-dependent reflection high-energy electron diffraction (RHEED) intensity profile of the specular spot during BTO growth under (b) $P_{O2} = 150$ mTorr, (d) $P_{O2} = 5$ mTorr, and (f) O$_2$/Ar mixed atmosphere. Note that the solid line in (f) indicates the clear layer-by-layer growth up to three oscillations while dashed line shows the intensity decay with further growth. The insets in (b), (d), and (f) indicate the RHEED diffraction patterns before and after BTO growth. The open arrows indicate the timing to stop laser pulses.

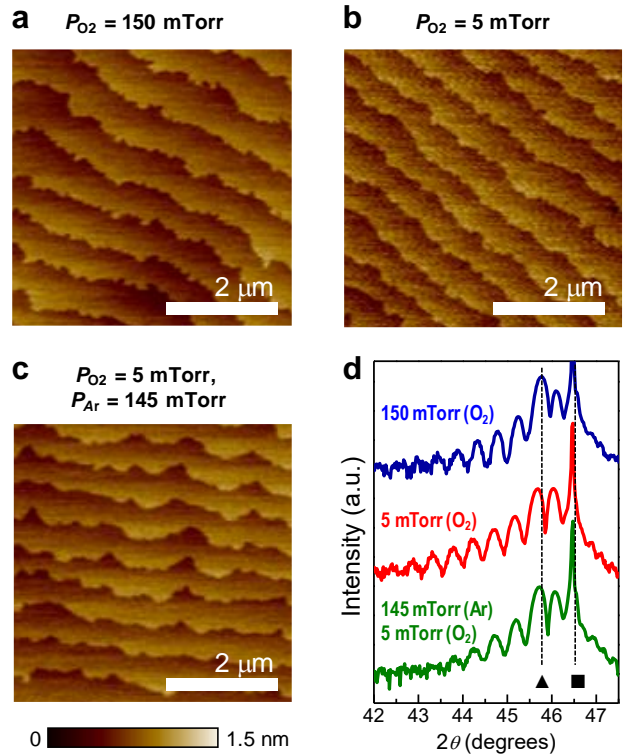

**Figure 2.** Structural characterizations of the SrRuO3 (SRO)/BTO/SRO heterostructures. (a-c) Atomic force microscopy (AFM) images of the SRO/BTO/SRO heterostructures with few-unit cells (uc)-thick BTO layer deposited at (a) $P_{O2}$ = 150 mTorr, (b) $P_{O2}$ = 5 mTorr, and (c) O$_2$/Ar mixed atmosphere. d) X-ray diffraction (XRD) $2\theta$-$\omega$ scans of the SRO/BTO/SRO heterostructures. The peak positions of SRO(220) and SrTiO$_3$(002) [STO(002)] peaks are labeled by solid triangles and squares, respectively.

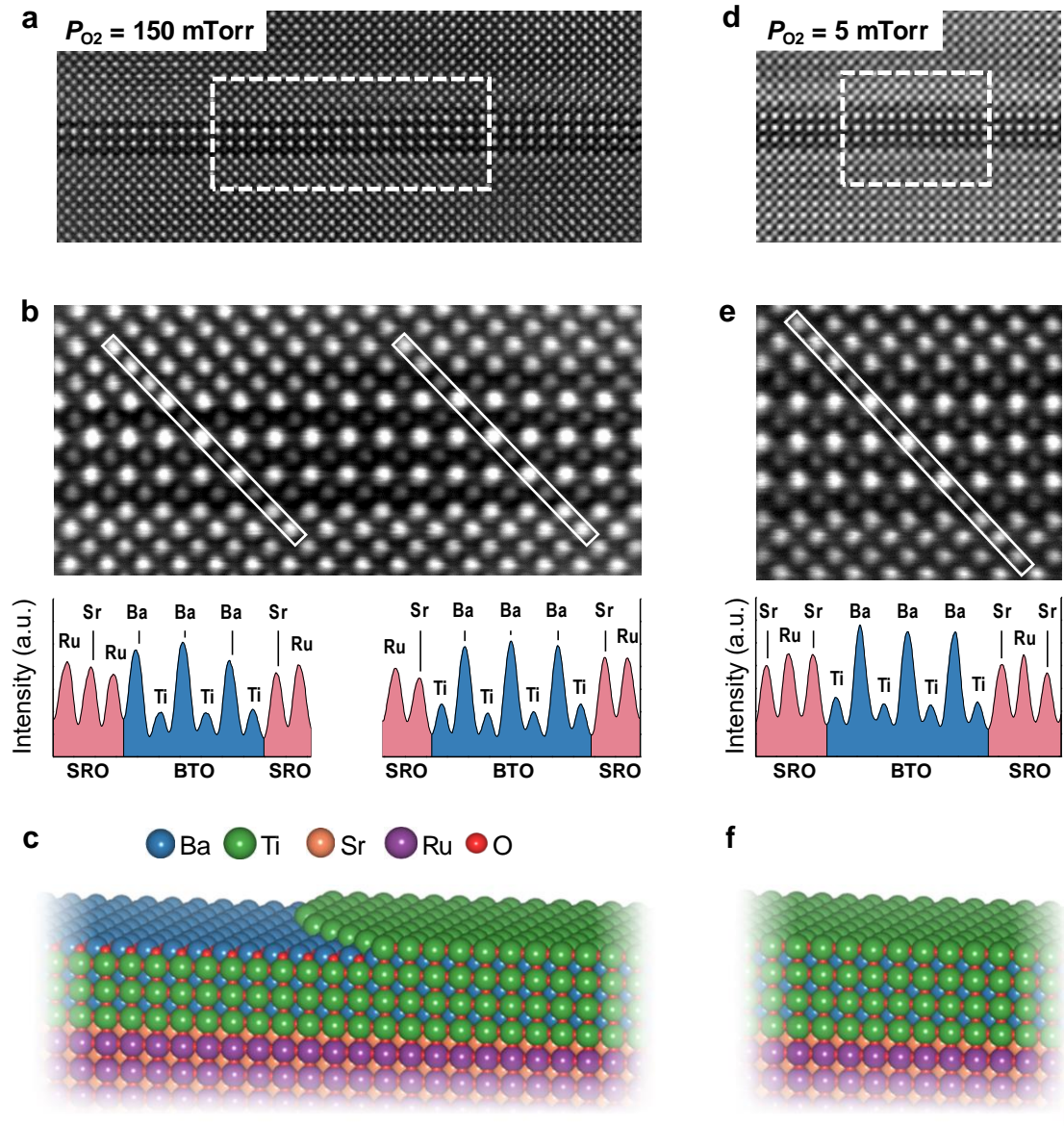

**Figure 3.** Interfacial structure of SRO/BTO/SRO heterostructures with BTO grown under pure oxygen atmospheres. (a) [(d)] High-angle annular dark field (HAADF) images of the SRO/BTO/SRO heterostructure with a BTO layer grown at $P_{O2}$ = 150 mTorr (5 mTorr). (b) [(e)] Magnified HAADF images of the pre-scanned area marked by the dashed box in (a) [(d)]. The HAADF intensity profiles along the solid boxes marked in (b) [(e)] are also shown. (c) [(f)] Schematic illustration of the BTO surface after the growth at $P_{O2}$ = 150 mTorr (5 mTorr). After BTO growth at $P_{O2}$ = 150 mTorr, the film surface (i.e., SRO/BTO interface) shows a heterogeneous termination sequence. In contrast, the as-grown BTO at $P_{O2}$ = 5 mTorr film surface shows a uniform SrO-TiO$_2$ termination sequence.

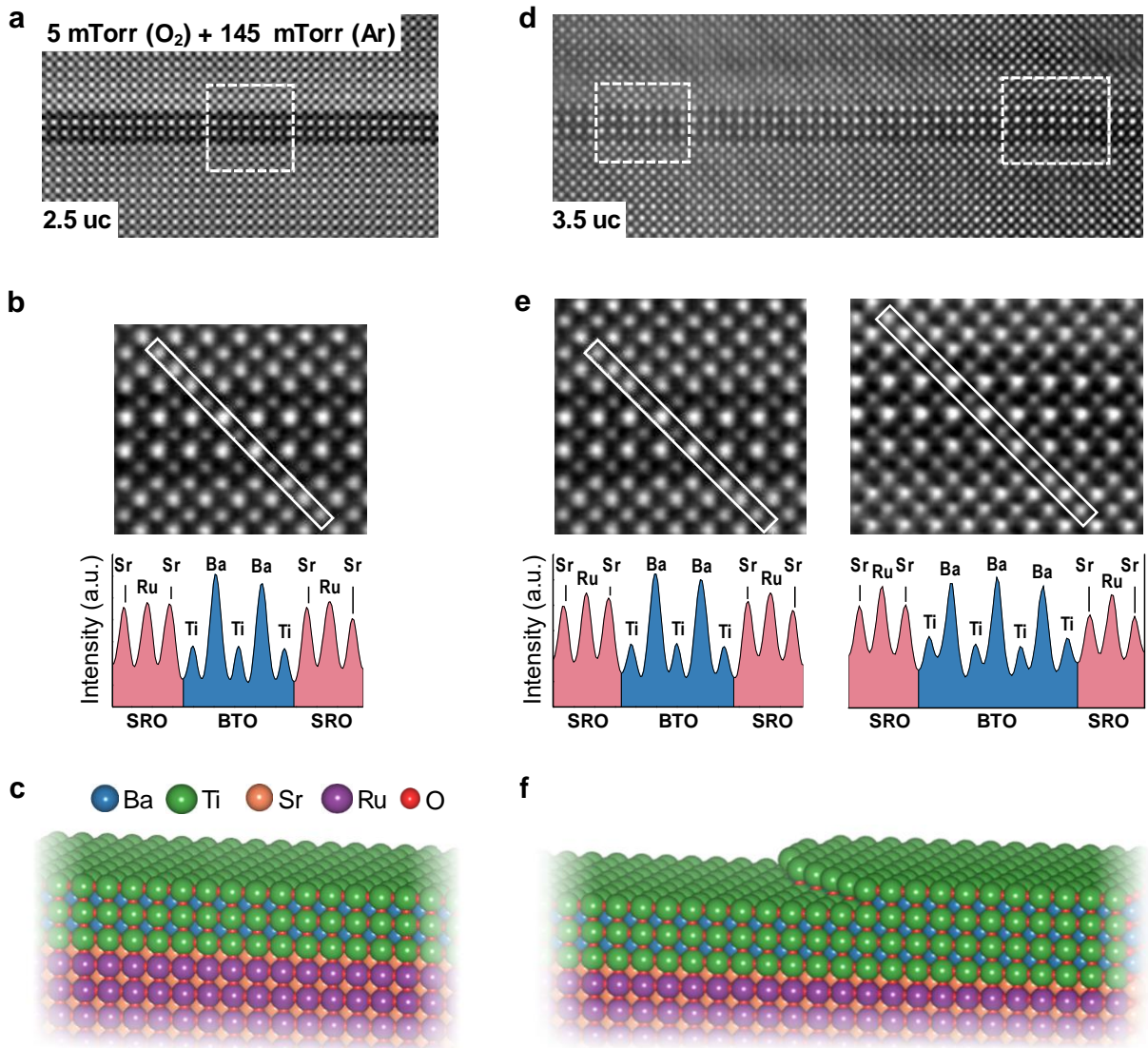

**Figure 4.** Interfacial structure of SRO/BTO/SRO heterostructures with BTO grown under an O$_2$/Ar mixed atmosphere. (a) [(d)] HAADF images of the SRO/BTO/SRO heterostructure, in which the BTO layer thickness is 2.5 uc (3.5 uc). (b) [(e)] Magnified HAADF images of the pre-scanned area marked by the dashed box in (a) [(d)]; HAADF intensity profiles along the solid boxes marked in (b) [(e)] are also shown. (c) [(f)] Schematic illustration of the 2.5-uc-thick (3.5-uc-thick) BTO film surface. For both 2.5- and 3.5-uc-thick BTO samples, the BTO surface (i.e., the top SRO/BTO interface) shows a uniform termination sequence.

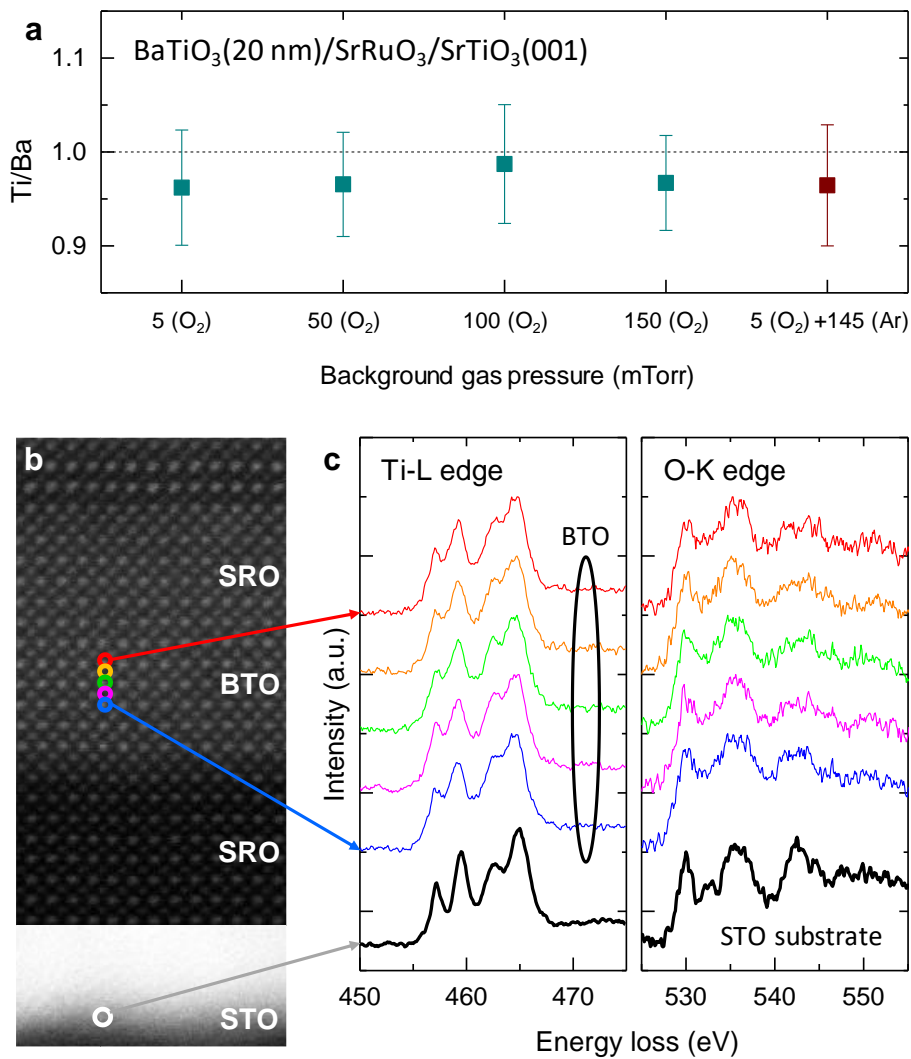

**Figure 5.** Characterization of the chemical stoichiometry of BTO films. (a) The cation Ti/Ba ratio of BTO films grown at different background atmospheres measured by energy-dispersive spectroscopy (EDS). The error bars indicate the standard deviations of Ti/Ba ratio measured from 20 locations randomly selected from one BTO sample. (b) HAADF images of the SRO/BTO/SRO heterostructure, in which the 2.5-uc-thick BTO layer was grown at $P_{O2}$ = 5 mTorr. (c) Electron energy-loss spectroscopy (EELS) at Ti-L and O-K edges measured from BTO layer and STO(001) substrate. The electron spot position is marked by the open circles and solid arrows in (b).

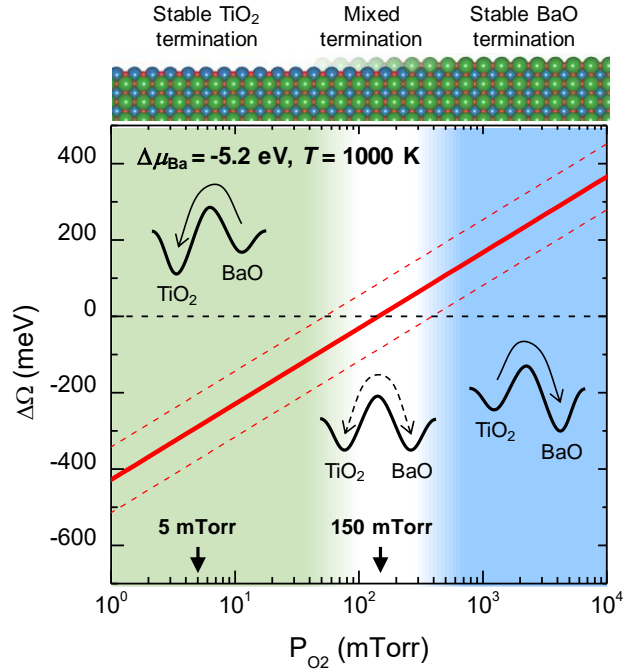

**Figure 6.** $P_{O2}$-dependent surface energy differences between TiO$_2$-terminated and BaO-terminated BTO ($\Delta\Omega = \Omega^{TiO2} - \Omega^{BaO}$). We set $\Delta\mu_{Ba} = -5.2$ eV and $T = 1000$ K. The dashed lines indicate the offset of $\Delta\Omega$ induced by the thermal energy at 1000 K (86 meV). The inset schematic diagram describes the possible energy landscapes of BTO surface structures (i.e., TiO$_2$ and BaO terminations), which evolve with $P_{O2}$. The arrows indicate the trend of stabilizing the specific surface structure.

**AUTHOR INFORMATION**

**Corresponding Author**

*E-mail: lingfei.wang@outlook.com (L.F.W.).

*E-mail: cshyoung@cau.ac.kr (S.H.C.).

*E-mail: twnoh@snu.ac.kr (T.W.N.).

**ACKNOWLEDGMENT**

This work was supported by the Institute for Basic Science in Kore (Grant No. IBS-R009-D1) and S.H.C. was supported by Basic Science Research Programs through the National Research Foundation of Korea (NRF-2015R1C1A1A01053163).

# Supporting Information for

"Oxygen Partial Pressure during Pulsed Laser Deposition: Deterministic Role on Thermodynamic Stability of Atomic Termination Sequence at SrRuO$_3$/BaTiO$_3$ Interface"

Yeong Jae Shin[†,‡], Lingfei Wang[†,‡,*], Yoonkoo Kim[§], Ho-Hyun Nahm[†,‡], Daesu Lee[†,‡], Jeong Rae Kim[†,‡], Sang Mo Yang[∥], Jong-Gul Yoon[#], Jin-Seok Chung[⊥], Miyoung Kim[§], Seo Hyoung Chang[¶,*], Tae Won Noh[†,‡,*]

[†]Center for Correlated Electron Systems, Institute for Basic Science (IBS), Seoul 08826, Republic of Korea

[‡]Department of Physics and Astronomy, Seoul National University, Seoul 08826, Republic of Korea

[§]Department of Materials Science and Engineering and Research Institute of Advanced Materials, Seoul National University, Seoul 08826, Republic of Korea

[∥]Department of Physics, Sookmyung Women's University, Seoul 04310, Republic of Korea

[#]Department of Physics, University of Suwon, Hwaseong, Gyunggi-do 18323, Republic of Korea

[⊥]Department of Physics, Soongsil University, Seoul 06978, Republic of Korea

[¶] Department of Physics, Chung-Ang University, Seoul 06978, Republic of Korea

*E-mail: lingfei.wang@outlook.com (L.F.W.).

*E-mail: cshyoung@cau.ac.kr (S.H.C.).

*E-mail: twnoh@snu.ac.kr (T.W.N.).

# I. RHEED monitoring of the SRO/BTO/SRO heterostructure growths.

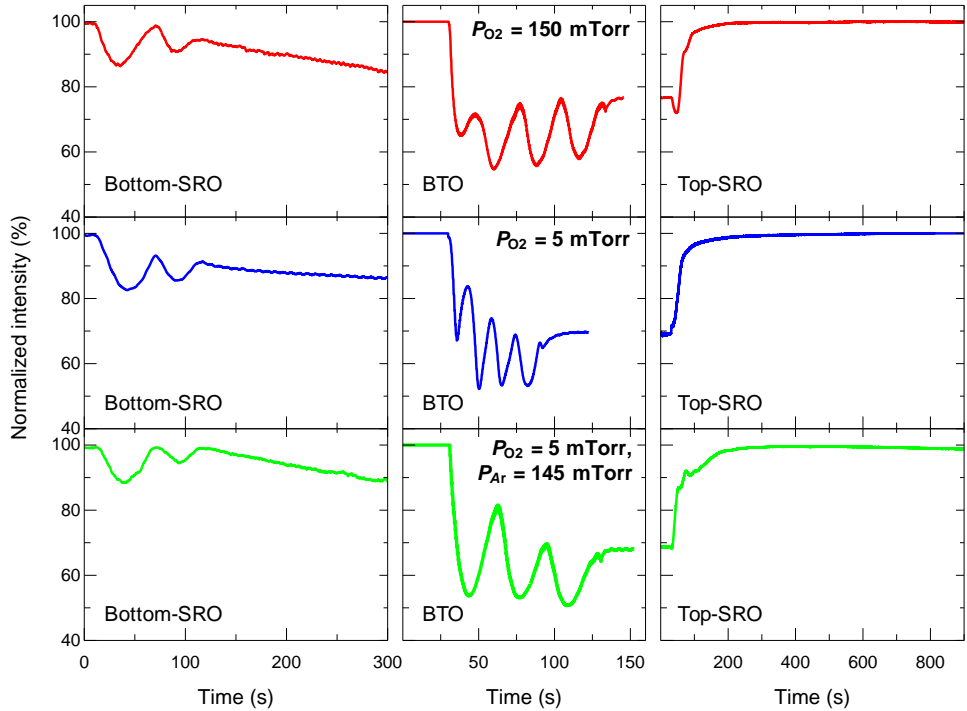

**Figure S1.** Time-dependent reflection high-energy electron diffraction (RHEED) intensity profiles of the specular spot during SrRuO$_3$/BaTiO$_3$/SrRuO$_3$ (SRO/BTO/SRO) growths under $P_{O2}$ = 150 mTorr (red), $P_{O2}$ = 5 mTorr (blue), and O$_2$/Ar mixed atmosphere (green).

The growths of SrRuO$_3$/BaTiO$_3$/SrRuO$_3$ (SRO/BTO/SRO) heterostructures were monitored by using *in situ* reflective high-energy electron diffraction (RHEED). **Figure S1** shows the intensity profiles of a specular spot of RHEED during the growths of bottom-SRO, BTO, and top-SRO layers. The BTO layers were grown under three total background gas pressure ($P_{total}$) and oxygen partial pressure ($P_{O2}$) conditions; 1) a pure oxygen atmosphere with $P_{total}$ = $P_{O2}$ = 5 mTorr (upper), 2) a pure oxygen atmosphere with $P_{total}$ = $P_{O2}$ = 150 mTorr (middle), and 3) an argon and oxygen (O$_2$/Ar) mixed atmosphere with $P_{O2}$ = 5 mTorr and $P_{total}$ = 150 mTorr (lower). The background atmospheres for bottom- and top-SRO layers were fixed to $P_{total}$ = $P_{O2}$ = 100 mTorr. The bottom SRO layers show a standard growth mode tradition from layer-by-layer mode to step-flow growth mode at first few layers.[1] Please note that the slight intensity decrement of bottom-SRO occurs because the electron beam was focused on the substrate surface and gets out of focus the more layers are grown. For the top SRO growths, the RHEED intensity profiles show a slight difference at the initial stage depending on the growth conditions, which may be related to the imperfect surface of BTO layer. However, all the RHEED intensity profiles of top-SRO layers saturated during the further growth, indicating a standard step-flow growth mode.

**II. Methodology for calculating relative BaTiO₃ surface stabilities**

The evolution of BaTiO₃ (BTO) surface stability can be described in terms of the surface Gibbs free energy for different surface structures and background atmospheres.[2,3] The surface Gibbs free energy of the $i$-th termination of BTO ($\Omega^i$) can be written as

$$\Omega^i = \Delta E^i - n_{Ba}\mu_{Ba} - n_{Ti}\mu_{Ti} - n_O\mu_O, \tag{S1}$$

where $\Delta E^i$ is the relative total energy of the heterostructure surface except for the fixed region that limits atomic relaxation as a bulk-like region per surface area; $n_{Ba}$, $n_{Ti}$, and $n_O$ denote the numbers of Ba, Ti, and O atoms in the surface per surface area, respectively; and $\mu_{Ba}$, $\mu_{Ti}$, and $\mu_O$ are their corresponding chemical potentials, respectively. For convenience, we measured the chemical potentials for each species with respect to their standard states at 0 K, for example, $\Delta\mu_a = \mu_a - E_a^{std}$. The reference state for oxygen is O₂ gas, such that $E_O^{std} = E_{O_2}^{gas}/2$. The reference states for cations are their respective bulk elemental phases. In addition, we can impose constraints on the chemical potentials of each atom species and the formation energy of bulk BTO, $\Delta G_f(BTO)$:

$$\Delta\mu_{Ba} + \Delta\mu_{Ti} + 3\Delta\mu_O = \Delta G_f(BTO). \tag{S2}$$

If we combine eq S1 and eq S2, we can express $\Omega^i$ with the chemical potentials of Ba and O as variable:

$$\Omega^i = \phi^i + \Gamma_{Ba}^i \Delta\mu_{Ba} + \Gamma_O^i \Delta\mu_O, \tag{S3}$$

where $\phi^i = \Delta E^i - n_{Ti}E_{BTO}^{bulk} + \Gamma_O E_O^{gas}/2 + \Gamma_{Ba}E_{Ba}^{bulk}$, $\Gamma_{Ba} = -(n_{Ba} - n_{Ti})$, and $\Gamma_O = -(n_O - 3n_{Ti})$. Note that all of these coefficients are intrinsic properties for each termination and independent of chemical environment. With the simple assumption of bulk-like (1 × 1) surface termination, the surface Gibbs free energies of the TiO$_2$- and BaO-terminated BTO surfaces can be written simply as

$$\Omega^{TiO_2} = \phi^{TiO_2} + \Delta\mu_{Ba} + \Delta\mu_O \text{ and} \tag{S4}$$

$$\Omega^{BaO} = \phi^{BaO} - \Delta\mu_{Ba} - \Delta\mu_O , \tag{S5}$$

respectively. eq S3 and eq S4 demonstrate that decreasing $\Delta\mu_O$ monotonically decreases the surface energy difference $\Omega^{TiO_2} - \Omega^{BaO}$, making the TiO$_2$-terminated BTO surface more stable. For quantitative investigations of the relative BTO surface Gibbs energy, we summarize the important thermodynamic parameters from our DFT calculations and previous experiments in Table S1.

Table S1: Thermodynamic parameters for obtaining the relative surface Gibbs free energy of BTO. The values are obtained from the generalized gradient approximation (GGA) with the on-site *U* Perdew–Burke–Ernzerhof (PBE) scheme and from the published thermodynamic data[4] at room temperature.

|  | Values | Note |
|---|---|---|
| $E_{Ba}^{bulk}$ | −1.87 eV | GGA-PBE |
| $E_{O_2}^{gas}/2$ | −4.71 eV | Exp. |
| $E_{BTO}^{bulk}$ | −11.81 eV | GGA-PBE |
| $\phi^{TiO_2} - \phi^{BaO}$ | 6.73 eV | GGA-PBE & Exp. |

## III. Ferroelectric polarization switching properties of SRO/BTO/SRO heterostructures

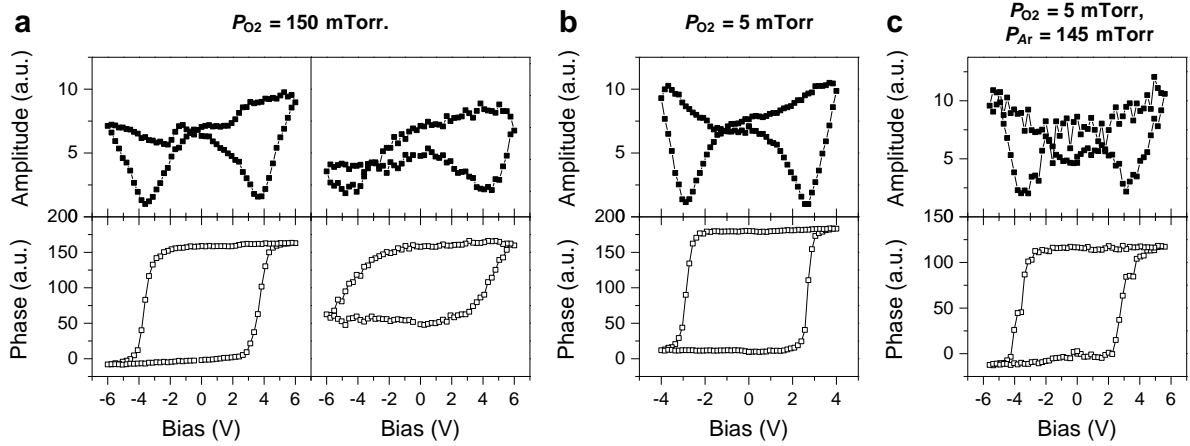

**Figure S2.** The amplitudes and phases from piezoresponse force microscopy (PFM) hysteresis loops of BTO films grown under (a) $P_{total} = P_{O2} = 150$ mTorr, (b) $P_{total} = P_{O2} = 5$ mTorr, and (c) $O_2$/Ar mixed atmosphere ($P_{total} = 150$ mTorr and $P_{O2} = 5$ mTorr).

The ferroelectric polarization switching properties of ultrathin BTO layers grown under three different background atmospheres were investigated by piezoelectric force microscopy (PFM). It was reported that the interface termination sequence at SRO/BTO interface affects ferroelectric properties. In particular, the formation of BaO-RuO$_2$ interface termination sequence at SRO/BTO generates pinned interface electric dipole which significantly disturbs the ferroelectric polarization stability in the ultrathin limit.[5] The interface termination sequence-dependent ferroelectric polarization properties can be found in our BTO films also.[6] Figure S2 shows the ferroelectric polarization switching of our BTO films measured by PFM. For $P_{O2} = 150$ mTorr case, both switchable (left, Figure S2a) and pinned ferroelectric polarization switching (right, Figure S2a) were found due to the coexistence of SrO-TiO$_2$ and BaO-RuO$_2$ interface termination sequences. On the other hand, for $P_{O2} = 5$ mTorr and $O_2$/Ar mixed atmosphere cases, clear ferroelectric polarization switching were observed due to uniform SrO-TiO$_2$ termination sequences at SRO/BTO interfaces (Figure S2b and S2c, respectively). Please note that the switchable ferroelectric polarization requires at least 3.5 uc thickness of BTO layer. The 2.5 u.c-thick BTO films do not exhibit ferroelectricity for both $P_{O2}$ = 5 mTorr and $O_2$/Ar mixed atmosphere cases.

**Supporting Information References**